\documentstyle[epsfig,psfig,subfigure,lscape,graphicx]{mn}
\makeatletter
\def\@cite#1#2{(\if@tempswa #2 \fi #1)}
\makeatother
\mathcode`\|="8000              
{\catcode`\|=\active \gdef|#1{_{\rm #1}}}

\begin{document}
\newcommand{\goodgap}{%
 \hspace{\subfigtopskip}%
 \hspace{\subfigbottomskip}}

\title[Instabilities in line-driven stellar winds]
{Non-spherical evolution of the line-driven wind instability}
\author[E.L. Gomez, R.J.R. Williams]
{E.L. Gomez\footnote{E-mail: edward.gomez@astro.cf.ac.uk}, R.J.R. Williams\footnote{Present address: AWE plc, Aldermaston, Berks; e-mail: 
robin.j.r.williams@awe.co.uk}\\ 
Department of Physics and Astronomy, Cardiff University, PO Box 913, CF24 3YB}

\maketitle
\begin{abstract}
In this paper, we study the structure and stability of line driven
winds using numerical hydrodynamic simulations.  We calculate the
radiation force from an explicit non-local solution of the radiation
transfer equation, rather than a Sobolev approximation, without
restricting the flow to one-dimensional symmetry.  We find that the
solutions which result have complex and highly variable structures,
including dense condensations which we compare to observed variable
absorption features in the spectra of early type stars.
\end{abstract}
\begin{keywords} hydrodynamics - instabilities - shock waves - 
Stars: mass-loss - winds, outflows
\end{keywords} 

\section{Introduction}

The winds of hot, luminous, early-type stars are composed of material
photoionized by the bright ultraviolet continuum of the star.  Lucy \&
Solomon (1970) proposed that resonance line scattering by metal ions
would provide a strong source of driving, to explain the high
mass-loss rates and flow velocities observed.  While a single
resonance line of gas at rest would only be driven by a small part of
the stellar spectrum, the many ($\approx 10^3$) resonance lines and
the variation of the line frequency which results from the Doppler
shift multiply the line force until it dominates other driving
mechanisms.

Detailed calculations can be made in which an accurate line-list is
used and radiative transfer is calculated using a photospheric
spectral energy density distribution, but the computational expense
required necessitates other approximations and the accuracy of the
results depends on the accuracy of the line-list used.  Castor, Abbott
and Klein~\shortcite{cak} (hereafter CAK) proposed an alternative
approach, where instead of using a detailed table of line opacities,
the force is calculated assuming a distribution function of the
numbers of lines present as a function of the line opacity, generally
specified as a power-law (indeed, the results of more detailed studies
are often characterised in terms of an effective power-law parameter).
CAK use the large-velocity gradient approximation (Sobolev 1960) to
allow the steady-state structures of the winds to be determined by
integration of a simple ordinary differential equation.

The Sobolev approximation is more suitable in a steady state model
than for a time-dependent simulation, since effects that evolve on a
small scale will not be adequently accommodated.  This issue was
addressed by Owocki, Castor and Rybicki~\shortcite{ocr} (hereafter
OCR) where the Sobolev approximation is replaced by an explicit
calculation of the optical depth.  In this paper, they use what they
term a pure-absorption approximation, in which photons which interact
with the gas are removed from the stellar spectrum.  While this
initially seems a strange approach to modelling a wind driven
predominantly by resonance scattering, the frequency redistribution
(in the rest frame of the star) achieved in the scattering process
means that it is a reasonable first approximation, and indeed the
accuracy of the results obtained have been confirmed by subsequent,
more detailed, studies by the same authors.

In the present paper, we apply the pure-absorption approximation of
OCR to study the local structure and stability of line-driven winds.
This work complements previous studies, which have generally either
used the non-Sobolev treatment pioneered by OCR \cite[and further
developed by]{owoc91,owocp96}, but been restricted to spherical
symmetry, or sacrificed the non-local treatment of the radiation field
to allow the flow to be studied in higher dimensionalities.  In common
limiting cases, our results agree with these other models to
reasonable accuracy.  When a non-local treatment of the radiation
transfer is combined with integration in a two-dimensional domain,
features result which are entirely novel. An initial effort to move
into the two dimensional domain was considered by Owocki (1999), where
they consider a multiple ray approach to the calculation of the line
driving.

The plan of this paper is as follows.  In Section 2 we discuss the
physical basis for a line driven from the Sobolev, CAK and OCR
formalisms. In Section 3 we discuss an approximate method for the
solution of the line driving force, the initial conditions used and
introduce three statistical parameters for use in analysing the
results. In Section 4 we introduce a 1D model, which can be compared
to previous calculations. The evolution of different
sizes of perturbation are considered, the data is statistically
analysed and the absorbtion spectrum is calculated. In
Section 5 the 1D model is extended to a 2D model and the resulting
structure is discussed. A summary is presented in Section 6.

\section{Radiative force and Hydrodynamics}

The absorption and scattering of radiation is the dominant process
which drives hot star winds.  The interaction of stellar photons with
wind particles accelerate the particles away from the star.  Treated
in full generality, the combined problem of hydrodynamics and
radiation transfer is a highly coupled integrodifferential system.
However, OCR demonstrated that a pure absorption model gave a
reasonable account of the local dynamics of wind instabilities in
spherical symmetry.  In this paper, we study the flow of gas in less
symmetrical situations, and adopt the pure absorption model as a
tractable first approximation.

The general expression for monochromatic optical depth in a wind is
\begin{equation}
\tau_{\nu}(r) = \int^{r}_{0} \rho(r')\kappa_{\nu}(r')dr',
\end{equation}
where $\rho(r)$ is the density of the gas at a distance $r$ from the
photosphere and $\kappa_{\nu}$ is the opacity at frequency $\nu$, per
unit mass of the wind material.  For a single absorption line, this
can be expressed in terms of a profile function,
\begin{equation}
\tau_{\nu}(r)= \int^{r}_{0} \rho\kappa|l 
	\phi\left(x(\nu)-\frac{v(r')}{v|{th}}\right)dr',
\end{equation}
where $x$ is the frequency displacement from line centre for gas at
rest with respect to the star, in units of the
line width, $\Delta\nu$,
\begin{equation}
x = \frac{\nu - \nu_0}{\Delta\nu},
\end{equation}
and $\Delta\nu = (v|{th}/c)\nu_0$ when the line width is dominated by
thermal broadening.  The lines are Doppler shifted by the local
velocity, $v(r)$ of the stellar wind.

The profile function, $\phi$, gives the relative likelihood that a
photon will be scattered as a function of the offset of the frequency
of the photon from line-centre in the local rest frame of the gas.  It
is given as a function of $x$, normalised so that
\begin{equation}
\int_{-\infty}^{\infty}\phi(x) dx = 1
\end{equation}
so $\kappa|l$ is the line opacity integrated over the line profile
divided by the the line width, $\Delta\nu$ (and so for a specific
state of the wind material, $\kappa|l v|{th}$ is independent of
$v|{th}$).  The line profile is determined from a combination of
intrinsic (quantum) line width, thermal motion and pressure
broadening.  For conditions which apply in a stellar wind, it may be
taken as Gaussian with a thermal width, although the Lorentzian wings
may be an important factor at the base of the wind \cite{poc90}.  The
variation of line profile between ions is small enough to be neglected
in the present work.  The role of photospheric Lorentz wings in wind
acceleration in B stars was investigated by Babel (1996).

Following OCR, the acceleration due to a line at a particular
frequency, $\nu_0$, and opacity $\kappa|l$, can be expressed as,
\begin{equation}
\label{g_nu}
g_{\nu_0 , \kappa|l} = \frac{\kappa|l v|{th}\nu_0 F_{\nu_0}}{c^2}
\int_{-\infty}^{\infty}dx \; l_0(x) e^{-\tau(x,r)}
\phi \left(x - \frac{v(r)}{v|{th}}\right) ,
\end{equation}
where the factor preceding the integral, 
\begin{equation}
g|{thin} = \frac{\kappa|l v|{th}\nu_0 F_{\nu_0}}{c^2}, \label{gthin} 
\end{equation}
is the line acceleration in the optically-thin limit.  The function
$l_0(x)$ is the photospheric profile of the line, so that
\begin{equation}
F_{\nu_0}(r) = F_{\nu_0} l_0(x) e^{-\tau(x,r)}
\end{equation}
gives the strength of the local radiation field at position $r$ and
offset $x\Delta \nu$ from the line centre $\nu_0$.

In a stellar wind, there are many ionic transitions with different
frequencies and line strengths.  Rather than treat this ensemble of
lines in detail, CAK developed a statistical approach based on a power
law distribution line number density as a function of opacity.  OCR
extended this model, by assuming that the line opacities were
distributed as
\begin{equation}
N(\kappa) = {1\over \kappa_0}\left(\kappa\over\kappa_0\right)^{\alpha-2}
\exp(-\kappa/\kappa|{max}),\label{ocrnk}
\end{equation}
including an exponential cutoff at large opacity.  Assuming that the
underlying continuum radiation is approximately constant over the line
absorption band, the radiation acceleration from the ensemble of lines
is given by
\begin{eqnarray}
\label{grad}
\nonumber g|{rad}(r) &=& \frac{N_0 F
\Gamma(\alpha)}{c}\left(\frac{v|{th}}{c}\right)^{\alpha} \\
&\times&\int^{\infty}_{-\infty}dx \frac{\phi(x -
v(r)/v|{th})}{(\eta(x,r) + 1/\kappa|{max} + \phi(x)/\sigma_c)^{\alpha}} 
\; ,
\end{eqnarray}
where $\eta(x,r)$ is the ratio of the optical depth to the opacity and
can be thought of as a profile weighted column density from the
photosphere to a point $r$ in the wind,
\begin{equation}
\label{eta}
\eta(x,r) = \frac{\tau(x,r)}{\kappa|l} =
\int^{r}_{r_*}dr'\rho(r')\phi\left(x - \frac{v(r')}{v|{th}}\right).
\end{equation}

As indicated by OCR, a useful way of determining the radiation force,
equation~(\ref{grad}), is to use differences of its integral, the
radiation pressure
\begin{eqnarray}
\label{pgrad}
\rho g|{rad} &=& \frac{d(P_{rad})}{dz} \\
\nonumber P|{rad} &=& - \frac{N_0 F \Gamma(\alpha)}{(1-\alpha)c}\left(\frac{v|{th}}{c}\right)^{\alpha}  \\
\label{prad}
&\times&\int_{-\infty}^{\infty}dx \left[ \eta(x,r)  + \frac{1}{\kappa|{max}} + \frac{\phi(x)}{\sigma_c} \right]^{1-\alpha} \; .
\end{eqnarray}
The above neglects the effects of re-emitted or scattered radiation.
One important such effect is the `line-drag' phenomenon, described by
Lucy (1984).  This can prevent small-scale velocity perturbations from
growing as a result of the anisotropic scattered radiation field.
While we neglect this effect in the present work, we discuss how our
methods may be extended to take account of its influence.

\subsection{The Sobolev approximations}

In the limit of narrow lines, the profile function reduces to a delta
function and the integral in equation~(\ref{eta}) can found
analytically, giving
\begin{equation}
\label{sob_tau}
\tau_\nu = \sum_{r_i} \kappa|l v|{th}\rho(r)  
	\left\vert\frac{dv}{dr}\right\vert^{-1},
\end{equation}
where the sum is over the positions, $r_i$, at which the line of
interest is Doppler shifted to frequency $\nu$.  This will only occur
once in a monotonically accelerating wind with a single line, and is
often assumed to only occur once in simple treatments of more general
flows.  This is known as the Sobolev, or large velocity gradient,
approximation.

The local radiation driving for a single optically thick line is then
given by
\begin{equation}
g|{Sob} = \left(\frac{\nu F_{\nu}}{c v|{th}
\rho}\right)\left\vert\frac{dv}{dr}\right\vert.\label{singlesob}
\end{equation}
For an ensemble of lines obeying the distribution
function~(\ref{ocrnk}), OCR find
\begin{equation}
g|{Sob} = {N_0F\Gamma(\alpha)\over(1-\alpha)c}
\left({1\over\rho c}\left\vert dv\over dr\right\vert\right)^\alpha
\left[(1+\tau|{max})^{1-\alpha}-1\over\tau|{max}^{1-\alpha}\right],
\label{ensemblesob}
\end{equation}
where
\begin{equation}
\tau|{max} \equiv {\rho\kappa|{max}v|{th}\over \vert dv/dr\vert},
\label{tauensemblesob}
\end{equation}
which gives the CAK acceleration~$g|{CAK}$ in the limit
$\kappa|{max}\to\infty$.

The ratio of the thermal velocity to the velocity gradient,
$\ell|{Sob} = v|{th}/\vert dv/dr\vert$, is called the Sobolev length.
This is the length scale over which the flow velocity increases by a
thermal velocity.  The Sobolev approximation is invalid when velocity
gradients are small and cannot be used to study the dynamics of
processes with length scales smaller than the Sobolev length.

\section{Numerical method}

We performed numerical simulations of line-driven winds using the code
\textsc{vh-1} (Blondin 1990).  This code uses the Piecewise Parabolic
Method (PPM) of Colella \& Woodward~\shortcite{colwo84} to solve the
equations of hydrodynamic grid, using Lagrangian coordinates to
advance the solution, which is then re-mapped onto an Eulerian grid.

We added an additional force corresponding to line driving by a
radiation field. The derivation of this force term is described in
detail in the following section. The gas is assumed to be
isothermal. We treat the flow relatively far from the photosphere,
which allows us to use plane-parallel symmetry and to ignore finite
disc effects and spherical divergence, but requires that the wind
velocity is everywhere supersonic.

The scaling of the problem is kept arbitrary in the present paper,
since we are interested in the general behaviour of line-driven winds.
As a result, our results are not restricted to particular stellar
parameters.  The dimensionless parameters which must be specified
include the ratio of the thermal velocity to isothermal sound speed,
which we choose to be $v|{th}/c|s = 0.5$ \cite{poc90}. This is the
maximum value which produces a stable flow and is still physically
plausible.  Lower values of this ratio are inclined to create a less
steep and more unstable flow.

We take the boundary conditions to be constant velocity and density at
the inner boundary and free outflow at the outer boundary throughout,
with reflective boundary conditions at the sides of the
two-dimensional simulations.  The size of the numerical grid was kept
constant, $800$ cells for one-dimensional simulations and $800\times
200$ for two dimensions.

The value of the CAK power-law line-ensemble index is fixed at a
typical value of $\alpha=0.7$ (OCR).

\subsection{Optical depth calculation}

The calculation of the radiation force, equation~(\ref{prad}), for a
particular cell reduces to differencing the values of the radiation
pressure at the interfaces of the cell,
\begin{equation}
P^i_{rad} \propto \int^{x_1}_{x_2}dx \left[ \eta(x,r) 
+\frac{1}{\kappa|{max}} + \frac{\phi(x)}{\sigma_c} \right]^{1-\alpha},
\end{equation}
where $\eta(x,r)$ is the profile weighted column density which is the
amount of radiation absorbed by the wind up to the point $r$.  The
direct calculation of the double integral implicit in this equation
can dominate the time for a computation.

However, since the radiative driving force is dependent not on the
radiation pressure but rather on its gradient, a local approach can be
taken.  At any position the transmitted spectrum will only change over
a small range of offsets $x$ around the value for the local mean flow,
and the radiation driving will be obtained from the momentum of just
these absorbed photons.  The force calculation can therefore be
speeded up considerably by storing the local scaled optical depth
$\eta(x)$, and advancing it by a short-characteristics method.  Only
the value of the previous spectrum step, $\eta_{i-1}$, and the
increase across the current cell $\Delta \eta_{i}$ are necessary to
calculate the driving force in cell $i$,
\begin{equation}
{d P|{rad} \over dz} = \sum_{n=x_1}^{x_2} [(\eta_n +
1/\kappa|{max})^{1-\alpha} - (\eta_{n-1} +
1/\kappa|{max})^{1-\alpha}]\Delta x.
\end{equation}

Using the quantity
\begin{equation}
\eta(x,r)^\star = \eta(x,r)
+\frac{1}{\kappa|{max}} + \frac{\phi(x)}{\sigma_c}
\end{equation}
in our numerical calculations allows the photospheric profile and
opacity limit to be absorbed in the initial conditions.  In
fact, we also neglect the reversing layer term, since the physical
domain is assumed to be far away from the photosphere.

\subsection{Choice of profile function}

In the present paper, we approximate the profile function by a top hat
function.  This allows the calculation of the increase in column
density to further simplified, by using the analytic formulae for the
convolution of this function with the column density profile within
the grid cell,
\begin{equation}
h_i(v) = \left\{\begin{array}{ll}
        \rho_i \Delta x_i/(v|u- v|d) & v|d < v < v|u \\ 
		 0 & \mbox{otherwise,} \\
\end{array} \right.
\end{equation}
\begin{equation}
\Delta \eta_i = \phi_i \otimes h_i,
\end{equation}
where we have taken the density to be constant within grid cell and
the velocity vary linearly between limits $v|d$ and $v|u$ at its
edges.  While these distributions differ from the profiles assumed in
the hydrodynamical solution, we are free to assume this in the spirit
of operator splitting.  As we shall see, the results are sensitive to
the way in which the velocity limits for the cells are determined from
the cell-centred values in the hydrodynamical solution.

In reality, it would be more appropriate to take the line profile
function to be a thermal Doppler function or Voigt profile.  However,
our approximation should not have too dramatic an effect on the
results.  Owocki \& Rybicki (1984) compare the effect of different
forms of the profile function with analytical asymptotic solutions,
for the ratio of the perturbed line force, $\delta g$ to the perturbed
velocity, $\delta v$.  They operate the line driving on a sinusoidal
velocity perturbation of the form
\begin{equation}
\delta v(z) = \delta v_0 e^{ikz},
\end{equation}
where $\delta v_0$ is the velocity perturbation that propagates along
with the mean flow of the wind, $k$ is the wavenumber of the
perturbation, and $z$ is the Cartesian distance away from the
photosphere.  A useful quantity to determine is the ratio of the
perturbation of the acceleration to the apploed velocity perturbation,
$\delta g/\delta v$.  Owocki \& Rybicki (1984) show that if this
quantity has a positive real part, the flow is unstable.

When the optical depth within a Sobolev length is large, $\tau_0
=\rho_0\kappa|l \ell|{Sob}\gg 1$, a top hat profile gives an analytic
solution matching the asymptotic forms
\begin{equation}
\label{thickg}
\frac{\delta g}{\delta v} \approx \omega_0\frac{ik}{\chi_0 + ik},
\end{equation}
where $\chi_0 = \rho_0\kappa$ is the line strength in the mean flow,
and $\omega_0=g|{thin}/v|{th}$ is the growth rate of perturbations in
the short-wavelength limit, and the growth rate decreases to zero for
long-wavelength perturbations.

In the optically thin limit, $\tau_0 \ll 1$, this ratio becomes,
\begin{equation}
\label{thing}
\frac{\delta g}{\delta v} \sim \tau_0
\end{equation}
which is independent of the wavenumber and solely dependent on the
optical depth. This tends to zero making the perturbed acceleration
negligible.

By this perturbation analysis it is clear that while the flow feature
remains optically thick the flow will be stable.  This is shown in
equation~(\ref{thickg}) where only the phase of the initial
perturbation is affected by a change in the acceleration. In the limit
of short wavenumber perturbations, $\lambda \gg \ell|{Sob}$ the
perturbed force reduces to the Sobolev approximation $\delta g \sim
\delta v'$. The evolution of an optically thin perturbation,
equation~(\ref{thing}), is unstable since a change in the velocity
field produces a change in the acceleration. Since there is no
wavenumber dependence of the force in equation~(\ref{thing}) the
acceleration will continue until the perturbation becomes optically
thick.

The form of the profile function will have no effect on the radiation
driving force on scales larger than the Sobolev length.  As we find
(and as has been seen in previous work), the growth of instabilities
soon leads to smooth flow regions which are separated by strong shocks
and high-gradient rarefactions in which the Sobolev length will not be
resolved by the numerical grid.  In these latter regions,
discretisation errors in the hydrodynamical scheme will probably be at
least as important as errors due to our approximation to the line
profile function, and so it seems reasonable to use this
computationally convenient form in this first treatment.

While we have argued that the top hat-profile function is adequate for
use in this initial calculation, we will in future work extend this to
more realistic forms.  The generalisation may be achieved by
representing the profile using a sum of top hat functions, or with a
higher order approximation.

\begin{figure}
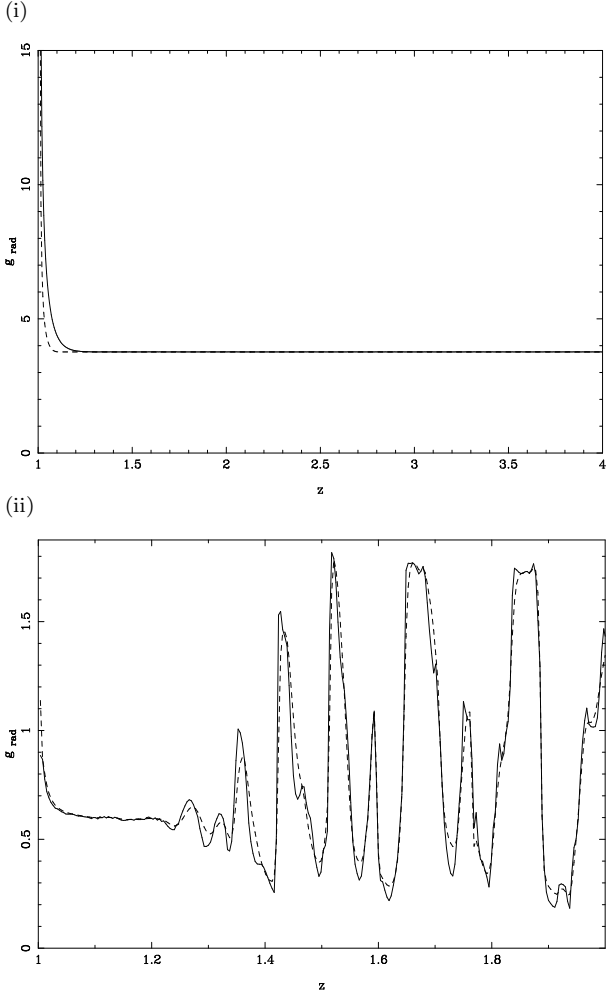

\begin{tabular}{l}
(i)\\
\psfig{file=prad_dv.25.ps,height=8.0cm,width=6.0cm,angle=-90}\\
(ii)\\
\psfig{file=prad_real.ps,height=8.0cm,width=6.0cm,angle=-90}
\end{tabular}
\caption{These plots show the acceleration from wind solutions using (i) constant density, $\rho=1$ and constant velocity gradient, $\Delta v/v|{th} = 0.25$, $\Delta x =0.05$ (ii) a driven flow which contains structure, including shocks (cf. \S~\ref{unpert_flow}). The Gaussian model curve (dashed) is overlaid on to the top-hat model curve (solid). The structure is very similar although the Gaussian example does produce a smoother acceleration profile for the case where there is much structure in the wind. The acceleration amplitude is given in code scaled units and are intended to illustrate the similarities between the two methods.}
\label{gauss_top}
\end{figure}

As a final word of justification to the use of a top-hat function in
this approximate wind solution we present a comparison of our top-hat
model with preliminary results using a Gaussian profile. We defer a
full comparison of the numerical effects of different profile shapes
to future work, so as not to detract from the main thrust of the
current paper. We compare results for the different profile functions
in two limits, (i) absorption due to a wind with constant velocity
gradient and constant density, (ii) absorption due to a data set,
representative of a wind solution containing small scale structure. As
can be seen from Fig.~\ref{gauss_top} there is little discernible
difference between the two approaches. These preliminary tests
indicate that the use of a top-hat function is not a serious over
simplification in the limits of this numerical implementation. We do
acknowledge that the Gaussian form of the profile function is a large
improvement over the top-hat and for future models where smaller
length and velocity scales are resolved,  it should be used. 

\subsection{Initial conditions}

We at first used a solution of the steady-state CAK equations to
provide initial conditions for our simulations.  However, it was found
that the change to a discrete grid resulted in significant disturances
in the subsequent flow solutions.  Instead, we ran the code without
perturbations and took the output data set as the starting point for
subsequent simulations.

We use constant velocity ($v=1$) and density ($\rho=1$) as the inner
boundary condition.  The radiation incident at the inside of the grid
was assumed to be a unabsorbed pure continuum.  Since the flow
considered here is supersonic at the outset, it is not consistent to
use a photospheric profile, as assumed by OCR.  As a result, there is
a narrow adjustment layer where the radiation field is first absorbed,
but this does not change the overall mass flux through the simulation.
It will, of course, be important to include the photospheric profile
in future work, when we include the geometric divergence of the flow
and calculate the development of the wind from the photosphere
outwards.

The values of the code units used throughout this work are shown in
Table \ref{data_vals}. Although arbitrary code units are used in the
numerical simulations, using typical physical values \cite{ocr} of
parameters (listed in Table \ref{data_vals}), the length units used
here are equivalent to stellar radii.
\begin{table}
\begin{tabular}{ll}
\hline
Code Parameter & Value \\ \hline
$v|{th}/c|s$ & 0.5 \\
$\alpha$ & 0.7 \\
$\kappa|{max}$ & $10^4$ \\ \hline
Physical Parameter & Value \\ \hline
$v|{th}$ & $40 {\rm km s^{-1}}$ \\
$\rho_0$ & $10^{-11} {\rm g cm^{-3}}$ \\
$N_0$ & $10.15 {\rm (cm^2g^{-1})^{1-\alpha}}$ \\
$F$ & $1.48 \times 10^{14} {\rm g s^{-3}}$ \\ \hline
\end{tabular}
\caption{Code and physical parameter values}
\label{data_vals}
\end{table}
\subsection{Statistical descriptors of the flow}

As suggested by Runacres \& Owocki~\shortcite{runao02}, the
time-dependent flow of the wind can usefully be characterized by three
statistical parameters: the clumping factor, velocity dispersion and
correlation function.  These give an indication of the mean level of
flow fluctuations at different points, and the form that these take.
In particular, the clumping factor will be unity and the velocity
dispersion will be zero in regions where the density varies little
with time, and significantly larger if the density has strong
transient features.  The correlation coefficient indicates if the
velocity and density are in (out) of phase, corresponding to a
dominance of forward (reverse) shocks in the flow, strong (anti)
correlation is shown when C $\rightarrow +(-)1$.

The clumping factor, velocity dispersion and correlation coefficient
are calculated using the relations
\begin{eqnarray}
\label{stats}
f_{{\rm cl},i} &=& 
\frac{\langle \rho_i^2 \rangle}{\langle \rho_i \rangle ^2}, \\
v_{{\rm disp},i} &=& \sqrt{\langle v_i^2 \rangle - \langle v_i \rangle^2}, \\
C_{v\log\rho,i} &=& \frac
{\langle v_i\log\rho_i\rangle - \langle v_i\rangle\langle\log\rho_i\rangle}
{v|{{\rm disp},i}\sqrt{\langle(\log\rho_i)^2\rangle 
- \langle\log\rho_i\rangle^2}},
\end{eqnarray}
where the angle brackets refer to a time average of flow quantities in
cell $i$; the parameters are functions of the position at which the
time average is calculated.  (The values were calculated only using
discrete samples in the evolution of the simulations, so in practice
we also extended the average over finite spatial regions to remove
spurious numerical noise.)

\section{One-dimensional simulations}

In this section, we describe one dimensional simulations using a
variety of forms for the radiation driving force.  We present these in
order to allow comparison of our results with those of previous
authors.  The simulations were performed both using a local driving
force, based on the CAK formalism, and using the pure absorption
approximation.

Previous studies, for example by Gayley et al.~\shortcite{gayig01} and
Proga et al.~\shortcite{proga98}, have applied local radiation driving
terms to flows in binary stellar systems and from the surface of
accretion discs.  We find here that the results so obtained are
crucially dependent on the manner in which the flow velocity gradient
is determined.

We then present the one-dimensional results of our treatment of the
pure absorption approximation.  The behaviour of an unperturbed flow
has been studied by many different authors (e.g. OCR, Runacres \&
Owocki~\shortcite{runao02}).  The basic stuctures of line driven flows
are rarefactions, steep reverse shocks as fast upstream material runs
into slow downstream material and de-shadowing of downstream gas,
where gas recieves an increase in acceleration from the radiation
field as it moves with sufficient velocity so that the incident
radiation no longer lies in the range that is absorbed by the upstream
patch.  We find good agreement with these results.  We then study the
development of flows with small and large perturbations.


\subsection{CAK force law}
\label{unstab_gr}
The radiation force may be calculated as a function of purely local
conditions using the CAK formalism.  As described above, this leads
to an acceleration term given by equation~(\ref{ensemblesob}) in the
limit $\tau|{max}\to\infty$, that is
\begin{equation}
g|{CAK} =
\frac{N_0 F \Gamma(\alpha)}{(1-\alpha)c}
\left(\frac{1}{c \rho} \left\vert\frac{dv}{dr} \right\vert \right)^{\alpha}.
\end{equation}
This has a milder dependence on velocity gradient than that given by
equation~(\ref{singlesob}), as a result of allowing for the effect of
locally optically-thin lines.

In order to implement the CAK driving force, a form of differencing
has to be chosen for the velocity gradient.  We investigate here two
extreme cases.  The upstream form is defined as
\footnote{This uses the convention that the inner boundary is at $n=0$
and the outer boundary is at $n=n|{max}$}
\begin{equation}
\label{usgrad}
\frac{\Delta v|{u}}{\Delta r} \rightarrow \frac{v_n-v_{n-1}}{r_n-r_{n-1}},
\end{equation}
and downstream form with
\begin{equation}
\label{dsgrad}
\frac{\Delta v|{d}}{\Delta r} \rightarrow \frac{v_{n+1}-v_n}{r_{n+1}-r_n}.
\end{equation}
The choice of differencing direction determines the direction in which
information is propagated by the radiation force term.  The upstream
calculation implies that the force depends on the gas properties in
the current and the previous cells.  In the downstream case, by
contrast, gas can `see' material in its shadow, which does not
correspond to the underlying physics of the radiation flow (at least
in the absence of scattering).

\begin{figure}
\rotatebox{270}{\psfig{file=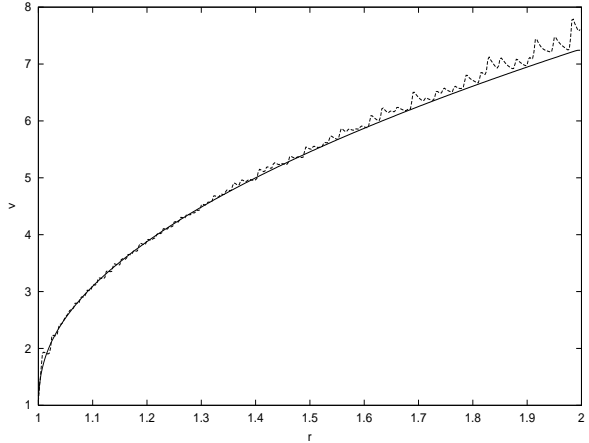,height=8.0cm,width=6.0cm}}
\caption{Instability in the wind produced by changing the direction of
velocity calculation. Dashed curve shows upstream gradient and solid
shows downstream. Both curves use the CAK radiation driving force
law.}
\label{unstab_grad}
\end{figure}

In Fig.~\ref{unstab_grad}, we compare the results of simulations using
these two forms.  We find that the downstream method,
equation~(\ref{dsgrad}), method is stable, and recovers the solution
of the steady-state CAK equations with good accuracy.  When the flow
is highly supersonic, the steady-state momentum equation becomes
\begin{equation}
v{dv\over dr} = g|{CAK},
\end{equation}
and since the mass flux $\Phi = \rho v$ is constant, $g|{CAK}\propto
(v\vert dv/dr\vert)^\alpha$.  As a result, the velocity increases as $v\propto
(r-r_0)^{1/2}$, which is a good approximation to the form of the solid
curve in Fig.~\ref{unstab_grad}.

However, the flow calculated using the upstream method,
equation~(\ref{usgrad}), is unstable.  If a stable steady flow is
obtained using the downstream method it will become unstable at every
point, when the method is changed to upstream calculation.  This
instability is not a short lived phenomenon but persists as long as
the upstream method is used.  This suggests that the instability is an
intrinsic property of the form of radiation driving law used, rather
than simple the result of initial flow perturbations and boundary
effects sweeping across the grid.

\label{abbott}

This behaviour may be understood in terms of the propagation of
`Abbott waves'.  Abbott (1980) treated the propagation of flow
perturbations in the Sobolev limit.  In one dimension, he found 2
radiative-acoustic wave modes moving at speeds
\begin{eqnarray}
v_- &=& v-{1\over 2}f|L'-\sqrt{\left({1\over 2}f|L'\right)^2+c|s^2} \\
v_+ &=& v-{1\over 2}f|L'+\sqrt{\left({1\over 2}f|L'\right)^2+c|s^2},
\end{eqnarray}
where $f|L'$ is the derivative of the radiation force with respect to
the velocity gradient.  The critical points in the steady-state CAK
equation occur where one or other of these mode velocities is zero.

The first of these radiative-acoustic modes is, however, somewhat
mysterious, as it moves {\it upstream}\/ through the flow at a speed
faster than the sound speed in the gas.  Owocki et al.~(1986)
demonstrated that Abbott's approach is justified only for purely
wave-like modes, in which the structure of the entire perturbation can
be analytically continued from any finite part of it.  This conclusion
remains somewhat controversial, although it is analogous to recently
observed laboratory phenomena such as the (apparently) superluminal
propagation of signals in samples of Bose-Einstein
condensates \cite{wang00}.  While information cannot propagate
upstream because of the purely downstream direction of photon
propagation, that does not prevent the phase and group velocities of
wave-like perturbations being directed upstream in an unstable flow.

Using the downstream velocity gradient in the calculation of the
Sobolev line force, however, allows information to travel upstream
through the grid, and as a result genuine modes with the properties of
Abbott waves arise in our numerical simulation.  The speed of
information propagation in these artificial waves is limited to $v <
-\Delta x/\Delta t$, which may be large if the time step is small on a
coarse grid.  The upstream velocity of these modes no longer {\it
requires}\/ the flow to be unstable, and in the outcome it is not.
This can be illustrated, for example, by observing that the
Abbott-like modes can allow the solution at the photosphere to adjust
to satisfy the conditions at the CAK critical point, and hence the
flow overall to relax to the steady-state CAK solution.

This stabilising feature of Abbott waves should, however, not be
dismissed out of hand as a numerical artifact, as scattered radiation
can lead to {\it some}\/ upstream propagation of information.  The
downstream differencing scheme can be thought of as combining the
(physical) upstream differencing scheme with a diffusion term
\begin{equation}
g \propto (\rho_{n+1}-\rho_n)/\Delta x 
	- \alpha (\rho_{n+1}-2\rho_n+\rho_{n-1})/(\Delta x)^2,
\end{equation} 
where $\alpha = \Delta x$ gives downstream differencing, and $\alpha =
\Delta x/2$ gives a centred gradient.  This diffusion term can then be
interpreted as a first-order approximation to the effect of photons
which scatter inwards at large radii and subsequently interact again
at smaller radii (the scattered photons which do not subsequently
interact are adequately treated by the absorption terms).

Specifically, the component $\rho_n-\rho_{n-1}$ corresponds to the
outward scattering force from photons which interact at radii beyond
the cell of interest: the downstream difference is allowable for these
photons as they subsequently move inwards, causing an {\it inward}
driving proportional to $\rho_{n+1}-\rho_n$.  We note that in this
simple model, flow stability only results when the inward force
resulting from scattering entirely cancels the direct force.  The
line-drag effect may be thought of as the result of just such a
stabilizing influence, although at a different level to that implicit
in fully downsteam gradient evaluation.

\subsection{OCR force law}

Initially the mean flow of the wind in one dimension was simulated
from the above model.  Although the model as described, is physically
incomplete it is useful to explore the most simple case of a
line-driven wind, i.e. unperturbed flow.  Unperturbed here means that
there is no extra perturbation inserted by hand, however the flow does
contain features that arise from its unstable nature in the optically
thin limit. This means that the flow is especially susceptible in the
early wind to instabilities.

As was shown previously for the CAK force, the choice of the direction
of velocity gradient calculation can effect the stability of the flow.
Since VH-1 stores variables as cell averages, the determination of
cell interface values for the velocity for use in the calculation of
the radiative acceleration requires the choice of an averaging method.
Two methods were used, a mixture of interpolation and extrapolation
from upstream information, and interpolation using downstream and
upstream information,

\begin{displaymath}
\textrm{mixture} 
\left\{ \begin{array}{ll}
v|u &= \frac{1}{2}(v_n + v_{n-1}) \vspace{0.1cm} \\
v|d & = \frac{3}{2}(v_n - \frac{1}{2}v_{n-1})
\end{array} \right.
\end{displaymath}
\begin{displaymath}
\textrm{interpolation} \left\{ \begin{array}{ll}
v|u &= \frac{1}{2}(v_n + v_{n-1}) \vspace{0.1cm} \\
v|d & = \frac{1}{2}(v_n + v_{n+1})
\end{array} \right.
\end{displaymath}
respectively.  The subscripts refer to upstream (u) and downstream (d)
values.  The mixture method is the most physically consistent
since it will not allow information to propagate upstream as a result
of the numerical method.  The interpolation method, since it uses both
upstream and downstream grid cells, does permit information
propagation upstream.

\subsubsection{Unperturbed flow}
\label{unpert_flow}
\begin{figure}
\epsfig{file=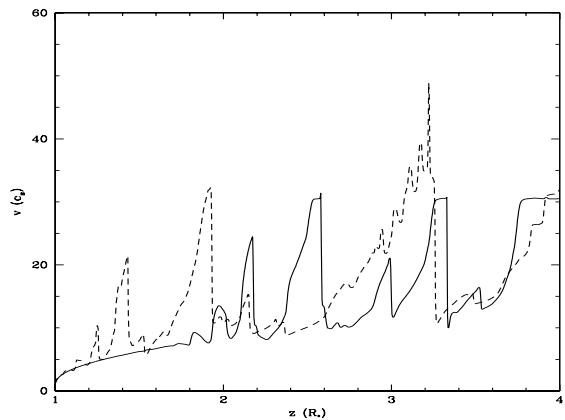,height=6cm,width=8cm}
\caption{A plot illustrating the difference in the direction which
velocities are calculated. Solid curve shows the evolution of the flow
using the interpolation method and the dashed curve the evolution using the
mixture method. The instabilities in the dashed curve appear
earlier in the flow compared with the interpolation method.}
\label{unstab_ocr}
\end{figure}

The results of these two methods of calculating the interface velocity
for a simulation of the structure of an unperturbed flow are presented
in Figure~\ref{unstab_ocr}.  It is apparent that while both methods
are unstable, this is true to a far greater extent for the
mixture method.

The instabilities are similar in form to the self-excited waves found
by OCR\@.  As in this previous paper, there is no clear mechanism
associated with their seeding, although it seems probable that
fluctuations caused by variations in the time step are amplified by
the line-driven flow instability.  In subsequent papers (Owocki 1991;
Owocki \& Puls 1996), Owocki \& his collaborators have adopted an
approach to the calculation of the scattered radiation, the Smooth
Source Function approach, which allows the wind to recover the CAK
solution when it is not perturbed at the photosphere.

However, as argued by Owocki \& Rybicki (1984), where the scattering
is small the wind should be intrinsically unstable throughout its
volume.  This is situation is similar to that often found in the study
of conventional turbulent flows, and various approaches have been
taken to treat it in that context.  In one approach, simulations are
regularized at the smallest resolvable lengthscales by a numerical
hyperviscosity, but an appropriate small-scale stochastic input should
be included {\it throughout the grid}\/ to model the effect of
sub-grid instabilities.  We here take an alternative approach, in the
spirit of large eddy simulations, allowing truncation errors to lead
to perturbations in the flow.  The exponential growth of the
instabilities and their saturation at non-linear amplitude should mean
that the overall properties of the flow are not strongly dependent on
the non-physical nature of their seeding.

While the dependence of updated flow values on the grid data is more
physically consistent in the mixture method, this method is also
highly unstable.  Some smoothing of the results is in fact desirable,
in order to ensure that structures continue to be resolved by the
numerical grid (this may be compared to the small smoothing of shocks
necessary to prevent post-shock ringing in Godunov-type schemes).  In
real stellar winds, the line-drag effect reduces the net effect of the
driving felt by these small wavelength disturbances at the base of the
wind.

The interpolation method will therefore be consistently used in the
remainder of this paper, and the flow shown by the solid line in
Figure~\ref{unstab_ocr} as the initial condition for all the remaining
simulations.  This has the additional benefit that the development of
finite perturbations may be studied in the relatively smooth region of
the solution between $z=1$ and $2$ without being confused by the
effects of rapidly-growing small-scale instabilities.

\paragraph{Statistical characterisation of the flow}
\begin{figure}
\begin{tabular}{l}
(a)\\
\psfig{file=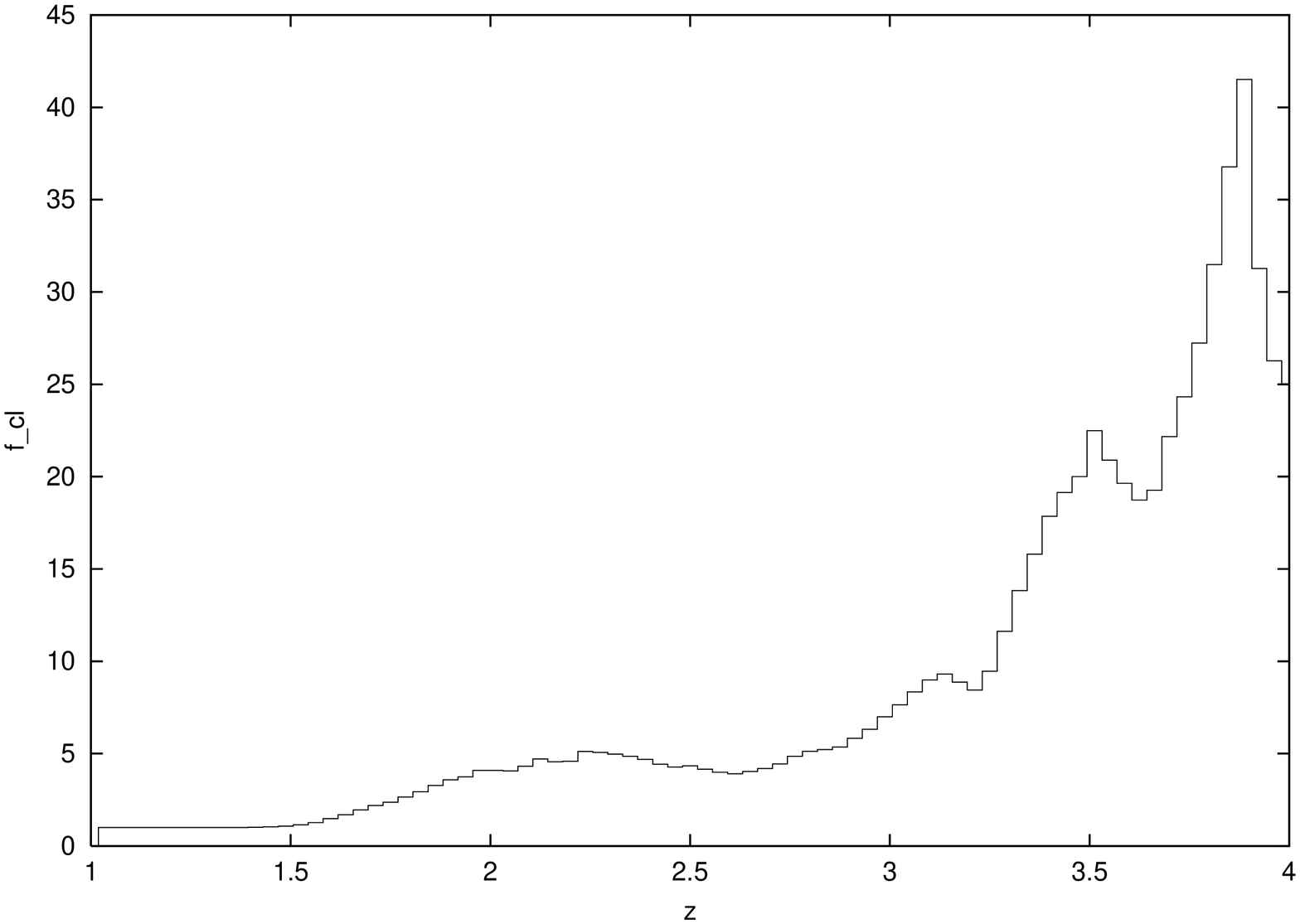,height=8.0cm,width=6.0cm,angle=-90}\\
(b)\\
\psfig{file=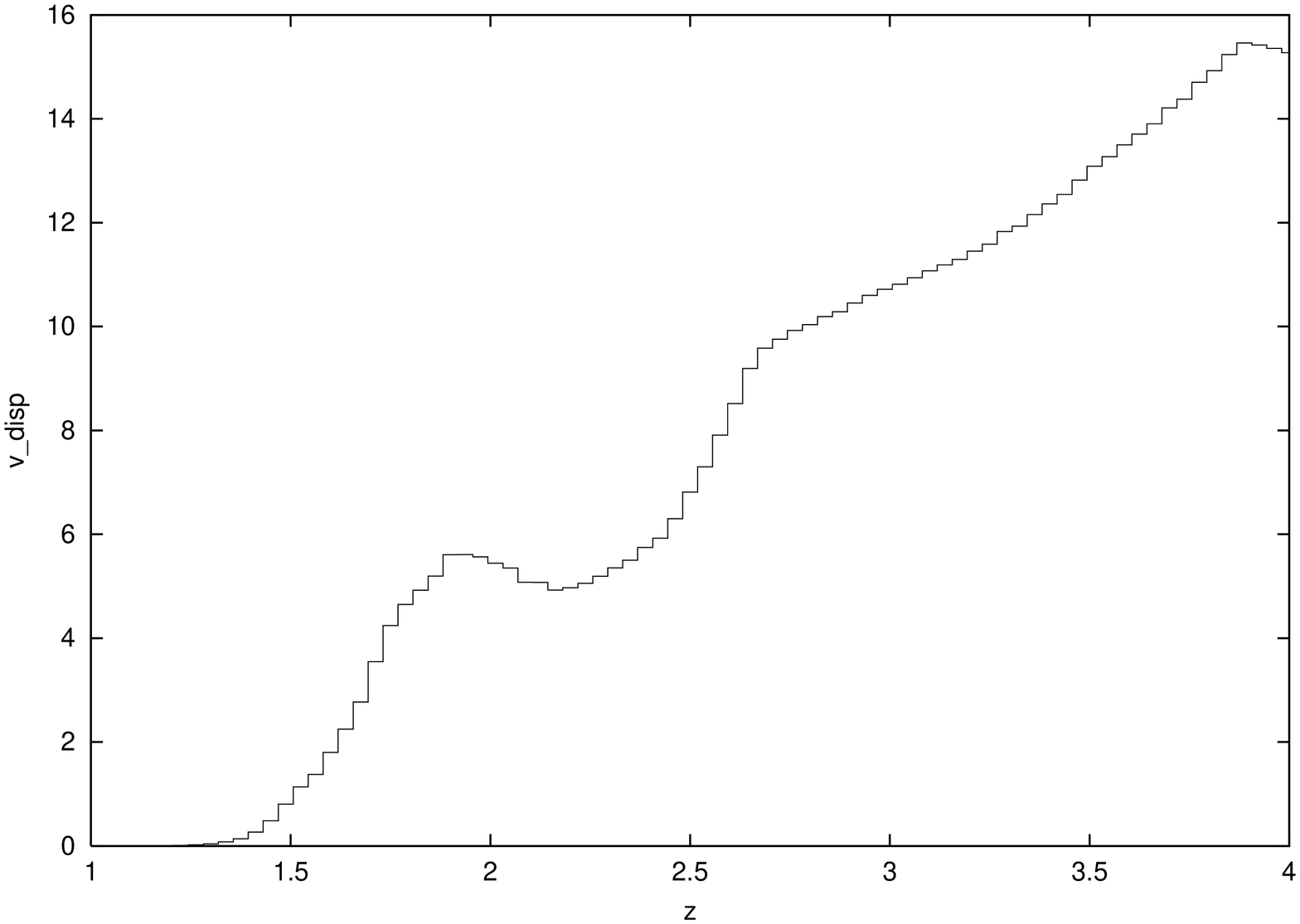,height=8.0cm,width=6.0cm,angle=-90}\\
(c)\\
\psfig{file=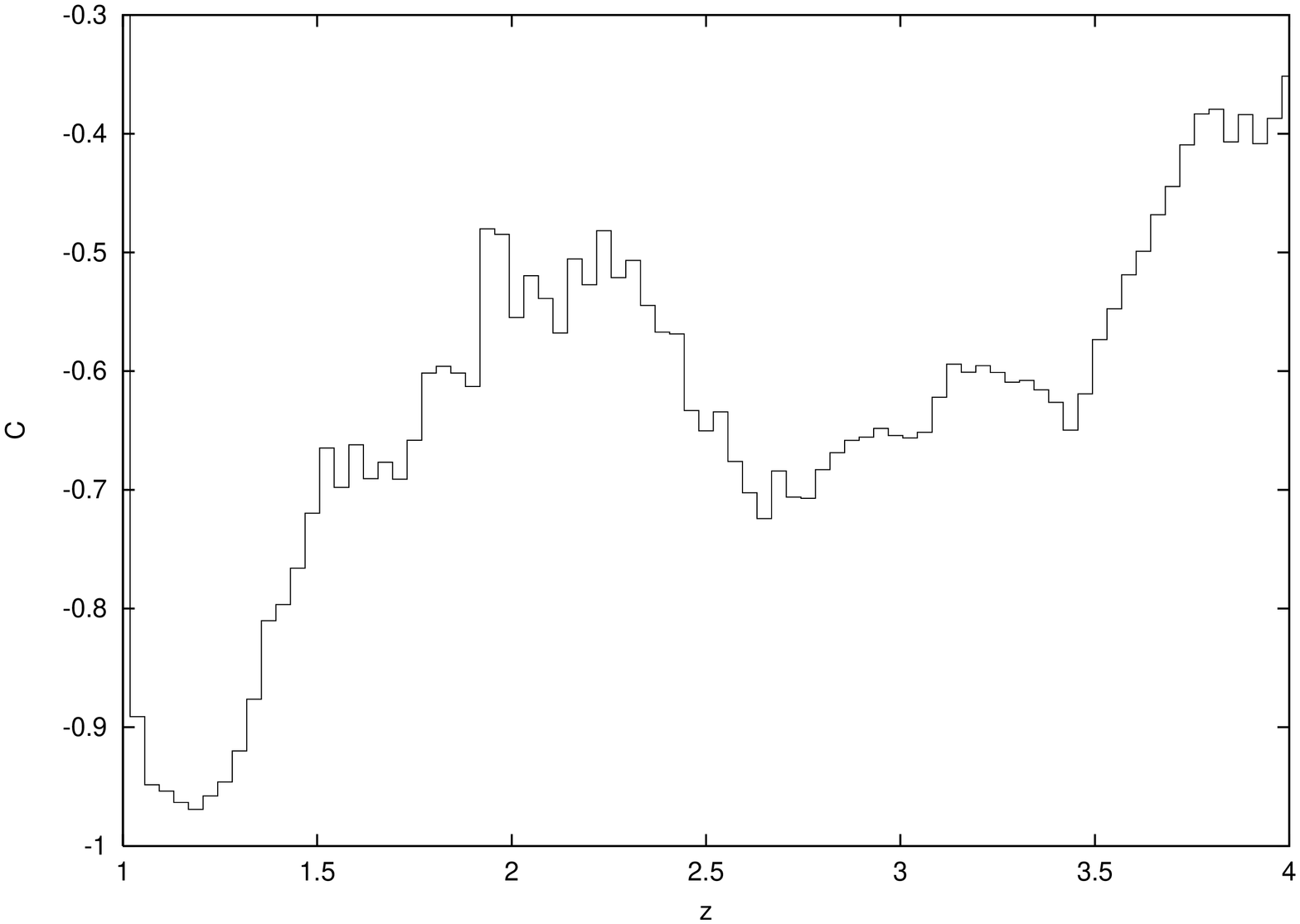,height=8.0cm,width=6.0cm,angle=-90}
\end{tabular}
\caption{The above graphs show the variation of each of the
statistical descriptors from equation~(\ref{stats}) with radial
distance.  The noise in the clumping factor is most evident, while the
other parameters are tightly constrained. The data has been binned
from 800 grid cells to 80 bins of equal size, to reduce the
scatter in the plots.  This scatter resulted from discrete features
which moved by many grid cells between the individual samples included
in the time averaging.}
\label{stats_graph}
\end{figure}

The overall structure of the flow variations may be characterized by
the statistical properties defined in equation~(\ref{stats}).  The
results for the above model, are shown in
Fig.~\ref{stats_graph}a-c.

As one might expect, the velocity dispersion, $v|{disp}$, is close to
zero at the base of the wind and the clumping factor, $f|{cl}$, is
approximately unity.  In this region there is little contribution to
the flow structure from the noise excited waves, so the wind is smooth
and has an almost constant velocity.

A minor peak in $f|{cl}$ is reached at $z=2$, the radius where the
noise waves begin to dominate the flow.  Upstream of this point there
are no dense shells but in the downstream flow the dense shells
steepen into shocks.  The flow contains many dense shells but it is
dominated by the accompanying rarefactions.  The noise in $f|{cl}$
beyind $z \approx 3 $ results from the extreme variability of the flow
in this regime, so an average over $100$ time units has still not
adequately sampled the behaviour of the flow in this region.
Nevertheless, it is clear that the amount of clumping and velocity
variation is increasing rapidly in the region.

Examination of $v|{disp}$ in Fig.~\ref{stats_graph}b shows that it too
reaches a minor peak at $z=1$.  Because this marks the point where the
diffuse clumps form into dense shells bounded by forward and reverse
shocks, $v|{disp}$ decreases as the smooth rarefied region begins to
dominate. This downturn is short lived as the rarefied regions broaden
and the shocked gas moves downstream.

The behaviour of $C$ is a little surprising but not totally
unexpected.  A flow dominated by reverse shocks will have $C \approx
-1$, since the velocity and density are anti-correlated.
Fig.~\ref{stats_graph}c shows this to be the case for the majority of
the wind studied here.  The initial portion of the wind moves sharply
from correlated at the inner boundary to anti-correlated after a short
distance, reaching a minimum at $z=1.2$.  This implies that reverse
shocks are the dominant structures, but it is unwise to ignore the
forward shocks.  Indeed, in the later stages of the flow each dense
shell is bounded by both an upstream reverse shock and a downstream
forward shock.  These forward shocks, although small in amplitude,
could begin to dominate the flow further downstream.

Comparing these results with Runacres \& Owocki (2002), shows good
agreement in all of the parameters except $C$.  They observe a
downstream value of $C \approx 0$ indicating that both forward and
reverse shocks are equally prevalent.  In the model here the reverse
shocks control the flow to a greater extent than the forward shocks,
borne out by the the slow rise, away from $C=-1$ to $C=-0.5$ at the
edge of the grid.


\paragraph{Absorption spectrum}

\begin{figure}
\psfig{file=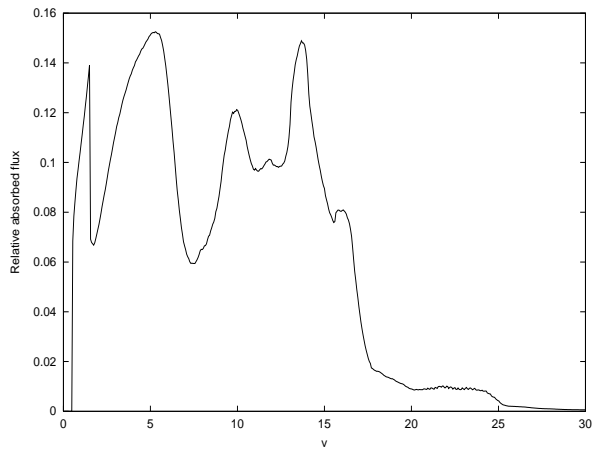,height=8cm,width=6cm,angle=-90}
\caption{A measure of the absorbed flux from the radiation field 
of the star. The high velocity end is incomplete and dominated by
noise.}
\label{abs1}
\end{figure}

An indication of the line absorption spectrum which would be found
from the flow can be found from the column density function,
$\eta(x)$, at the edge of the grid.  The high velocity part of the
profile is highly variable, so we will first look at the time average
of the structure.

In Fig.~\ref{abs1}, we show this for the above model.
Absorption profiles can be derived using
\begin{equation}
F(x) = \exp [-\kappa|{l} \eta(x)]
\end{equation}
where $\kappa|{l}$ is the opacity in a particular resonance line.
However, as our models are only of a restricted region of the wind,
such profiles would not be comparable in detail with observed P Cygni
profiles.

The results can nevertheless be compared to the theoretical
predictions of Abbott (1982).  Looking at the plots for $T|{eff} =
40,000{\rm\,K}$ and $50,000{\rm\,K}$ in this paper, they show
similarities to Fig.~\ref{abs1}, most notably between $v=0\mbox{--}7$.
The sudden drop in flux at $v=1.5$ where the steep decrease in density
stops and its behaviour becomes close to linear, representing the only
high density, low velocity region in the flow.  The smooth rise in
Fig.~\ref{abs1} is due to the smoothly increasing velocity and
decreasing density in the early part of the wind before noise begins
to dominate.  When a perturbation is applied to the base of the wind
the smooth region of $\eta$ in our model is disturbed and becomes
noisy showing the effect of perturbations on the spectrum.

\begin{figure}
\psfig{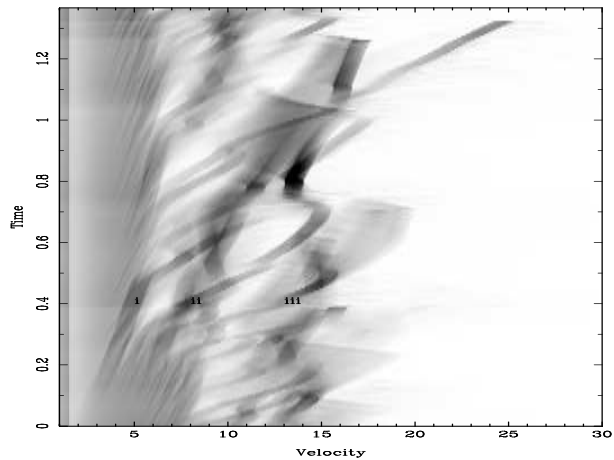}
\caption{A trail diagram plotting values of $\eta(x)$ as a function of
  time as a greyscale.  The broad lines in the plot indicate the path
  of optically thick shells -- the breadth of the lines is the width
  of the profile function.  Between $v =0 \mbox{--} 3.5$ the flow is
  smooth, instabilities grow exponentially in the range $v =3.5
  \mbox{--} 6$, so that most of the gas is concentrated in the
  ballistically-moving shells at larger radii.}
\label{trail_eta}
\end{figure}

The variation of $\eta(x)$ may be displayed in a trail plot,
Fig.~\ref{trail_eta}.  The development of dense shells in velocity
space is particularly clear in this plot.  At low velocities, the
trails are steep, with nearly constant gradient (i.e., the shells have
a low, constant acceleration).  As the shells reach a region where the
radiation field is no longer strongly absorbed by underlying gas, the
acceleration rapidly increses to a higher value, seen from the
shallower gradients of the trails.  Broad but gradually narrowing
regions of relatively smooth opacity appear at lower velocities than
some of the unresolved structures, corresponding to gas in the outer
part of these shells `surfing' on the enhanced radiation to the red
side of the saturated absorption band.  More complex structures
develop, as shells interact. 

Care must be taken when interpreting information from such a trail
plot, since flow structures with the same velocity can appear at
different places in the wind. This can make the origin of each trail
ambiguous. Nevertheless, a trail that remains well defined for a
significant time is the result of a single optically thick feature,
since since the combination of differently spaced features would not
have the same stability.

\begin{figure}
\psfig{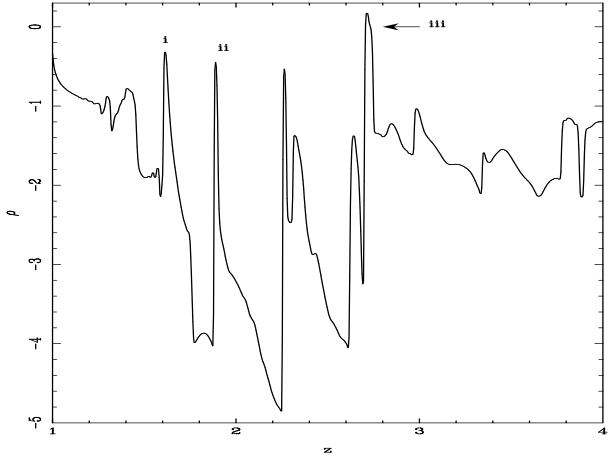}
\caption{This plot of $\rho$ against $z$ is taken at $t=0.414$. 
The dense shells labelled with I, II, III match those whose histories
are shown in the absorption trail plot, Fig.\ref{trail_eta}. }
\label{dens_trail}
\end{figure}

The points marked I, II, III have been identified with dense shells on
the density plot Fig. \ref{dens_trail}, which is taken at $t=0.414$. In
this case the three marked trails are the dense shells which stand out
against the flow as being the most optically thick. It is clear from
Figs. \ref{trail_eta} and \ref{dens_trail} that although the shell
marked I is not the densest shell, it has the highest optical depth
of the shells at this time. This is evident from its steep
evolutionary path, indicating that the region accelerates slowly and
as it recieves less driving.

 The point marked as III in Fig. \ref{trail_eta} indicates where the
 evolution of a dense lump changes from optically thin evolution to
 optically thick. The difference in the evolution is visible in two
 ways; the dark, steeper trail shows a marked difference to the
 earlier light, shallow trail of the optically thin
 evolution. This transition occurs because of minor shells colliding
 with the perturbation shell. As faster shells merge with the
 perturbation from upstream, the perturbation sweeps up slower
 shells downstream of it. These interactions increase the density of
 the shell and alter the optical properties making it optically
 thicker. As this behaviour continues the shell slows and at $t=0.485$
 it merges with another shell and its once distinctive optical depth
 becomes muddled with that of a shell further upstream which is
 forming out of the noise. The perturbation shell does not disappear
 but from a spectral perspective it is indistinguishable from other
 parts of the flow, whilst remaining the densest shell in the region
 of flow under consideration.


\subsubsection{Optically thin perturbation}

\label{1dpert}
\begin{figure}
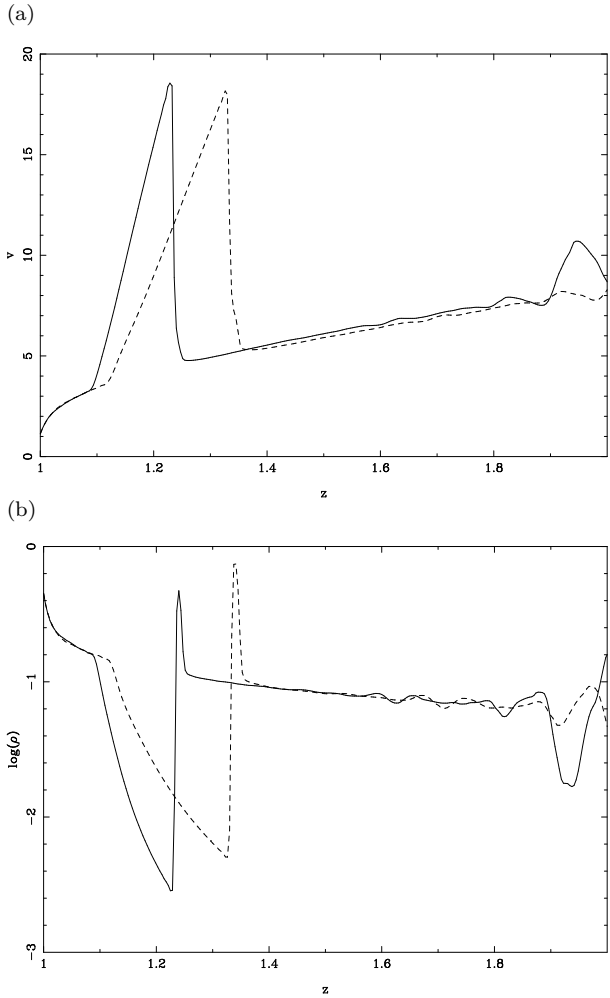

\begin{tabular}{l}
(a)\\
\psfig{file=thin_v1.ps,height=8.0cm,width=6.0cm,angle=-90}\\
(b)\\
\psfig{file=thin_r1.ps,height=8.0cm,width=6.0cm,angle=-90}
\end{tabular}
\caption{Velocity (a) and density (b) plots for flow with an optically
thin perturbation added.  Data from two different times have been
overlayed: $t= 0.005$ (solid), $0.01$ (dashed). The prominent features
of the plots are the advection of the optically thin perturbation in
the upstream portion of the wind. }
\label{thinv1}
\end{figure}

\begin{figure}
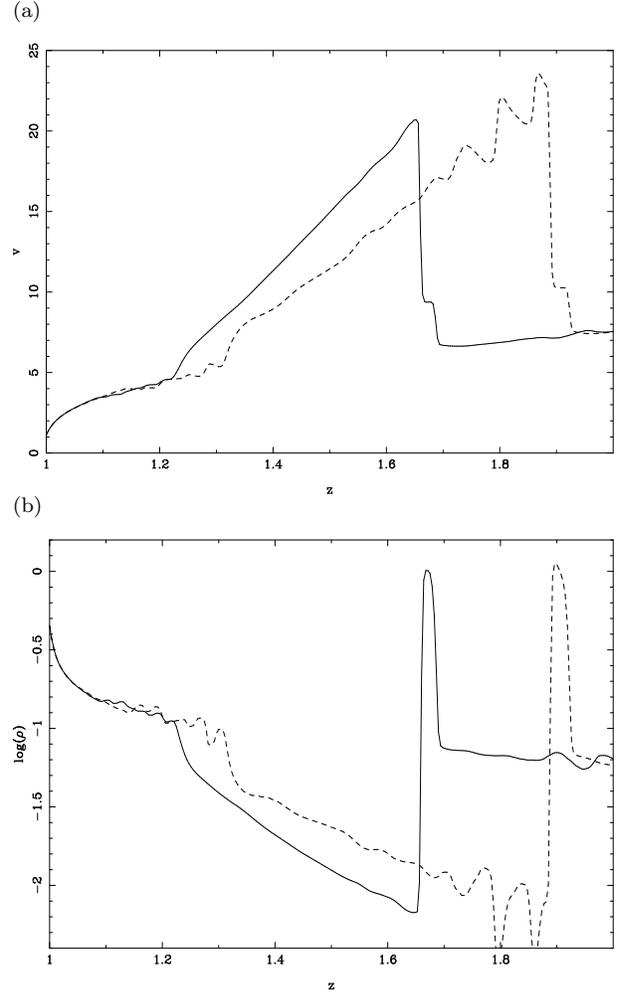

\begin{tabular}{l}
(a)\\
\psfig{file=thin_v2.ps,height=8.0cm,width=6.0cm,angle=-90}\\
(b)\\
\psfig{file=thin_r2.ps,height=8.0cm,width=6.0cm,angle=-90}
\end{tabular}
\caption{Velocity (a) and density (b) plots for the simulation shown in 
Fig.~\protect\ref{thinv2} at later times, $t= 0.016$ (solid), $0.021$
(dashed). The advected material now becomes unstable and begins to
break into dense shells. }
\label{thinv2}
\end{figure}

A perturbation is applied to the base of the wind which is 10 grid
elements thick ($\Delta z=0.0375$).  The amplitude of the
perturbation is a $v|p = 10 \times v_0$ in velocity, and the
change in density is $\rho|p = \rho_0/10$, the subscript $0$ refers
to background flow values; within this region, the
velocity was increased to 10 times its previous value, and the density
reduced by the same amount factor.  These values were chosen to represent
a typical short-wavelength ($1/k \ll \ell|{Sob}$) perturbation, to
compare with the analytical work from Owocki \& Rybicki (1984),
discussed above.

In this limit the evolution of the perturbation should follow
equation~(\ref{thing}).  The gas in the perturbation experiences a
constant acceleration, $\sim \tau_0 \delta v$.  In Fig.~\ref{thinv1}
the behaviour of the perturbation appears to be stable, as the gas in
the perturbation is swept up into a dense shell which advects slowly
across the grid.  Behind this shell, the constant velocity gradient is
characteristic of a free rarefaction.  

Immediately ahead of the shell, the flow remains smoother than
was the case in the unperturbed flow.  It seems that the strong shock
at the outer edge of the dense reduces the fetch available to the
rapidly moving unstable outward radiative-acoustic waves, and so
limits their amplitude in this region.

As the velocity gradient in the rarefaction decreases, instabilities
form at the edges of the rarefaction, as can be seen in
Fig.~\ref{thinv2}.  Behind the dense shell, subsidiary shells form,
while the smooth expansion of the gas upstream of the edge of the
rarefaction also becomes highly disturbed.  The optically thin
subsidiary shells gather momentum from line driving and run up the
diffuse wake of perturbation until they collide with the shock at the
leading edge of the dense shell.  This merging of shells is similar to
that which occurs in the outer wind.

The dense shell is not subject to the formation of instabilities and
is gently advected across the grid.  This supports the asymptotic
form, equation~(\ref{thickg}) which implies that this feature is
stable for small wavelength perturbations: any change in the radiation
force will only affect the phase of the perturbation, or the
derivative of the velocity.  The coupling of the radiation field with
the dense shell is very weak which is evident in its unchanging
amplitude, indicating that it is largely unaided by the driving force.

\begin{figure}
\psfig{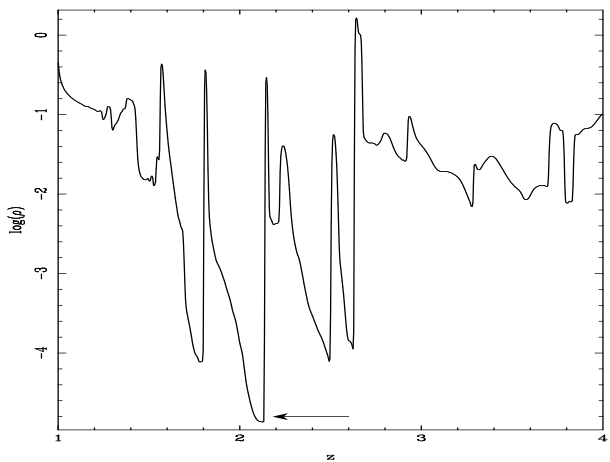}
\caption{A plot of the flow at $t = 0.028$, where reverse shock structure 
can be observed forming in the unstable wake of the initial
perturbation (indicated by the arrow)}
\label{r_shock}
\end{figure}

Shock structure begins to develop in the flow as an optically thin
flow encounters a dense shell.  A perfect example of a reverse shock
can be seen in Fig.~\ref{r_shock}, a dense shell with a smoothly
rarefied region preceding it. This figure also shows other shocks
forming and interacting with the surrounding dense regions.

\section{Numerical evolution of perturbations in two dimensions}

The non-Sobolev approach described in the present paper can be
naturally applied to multidimensional flows.  In the present paper, we
will concentrate on the simplest case, of the propagation of a single
ray through plane-parallel flow.  This is equivalent to studying a
very narrow slice of the wind at a reasonable distance from the
photosphere.  

This has the obvious advantage that only a single ray calculation need
be performed, parallel to one axis of the computational grid.  It may
be argued that this approach is little more than a series of
one-dimensional simulations in separate but contiguous channels.
However, while there is no tangential component to the radiation
driving, we will see that the flows which are driven in this direction
have a substantial effect on the structure of the wind.

The flows are studied in a similar way to the one-dimensional cases
presented above.  Perturbations are made to velocity field, added
close to the base of the wind.  A variety of widths and amplitudes of
perturbation were modelled, but in the end the evolution of the flow
always returns to structure of a similar form, so we will present here
a small selection of typical results.

An initial two dimensional data set was obtained by filling each row
of the simulation with identical data, derived from the unperturbed
one dimensional run shown in Figure~\ref{unstab_ocr}.  Running the
model without further perturbation, its development is similar
initially to one-dimensional runs.  However, as the simulation
progresses these structures fragment in the tangential direction.  It
is to be presumed that these instabilities are seeded by small
rounding errors in the underlying hydrodynamical scheme.  Certainly
the fragmentation initially appears nearly periodic.

The growth of these perturbations of the dense noise-excited shells
may be attributed to Richtmeyer-Meshkov instability (a particular case
of Rayleigh-Taylor instability which occurs when a shock interacts
with a density discontinuity).  As structure evolves from these
instabilities, it has a very definite effect on the downstream flow.
The result is a clumpy, highly structured flow as can be seen in
Fig.~\ref{clumpy}.

If the gas is diffuse the instabilities do not evolve to macroscopic
structure, because the strong coupling with the driving washes out the
structure.  Dense shells that have become optically thick, expand from
gas pressure alone.  This provides mechanisms by which both dense and
diffuse shells can expand.  Shocks are formed where the edges two such
expanding shells collide, for example at (ii) in Fig.~\ref{2d_wide1}b.
Depending on the expansion velocities of the shells involved,
hydrodynamical instabilities may develop at the interface.  The
development of instabilities in the flow can fragment the resulting
shell into multiply bullets and diffuse bubbles.

To study the development of these structures in more detail, we will
now investigate the development of flows in which an explicit
perturbation is included in the initial conditions.

\begin{figure*}
\caption{The evolution of the flow when no perturbation is added 
. The structure develops as RM instabilities become a dominant
feature in the flow. There is a complicated hydrodynamical interaction
between shadowing, de-shadowing, and gas pressure expansion. The
regular features result from the RM instabilities, however edge
effects begin to arise as well as a high level of unresolved structure
which disturbs the flow in an amorphous fashion. The axes  measure the same distance code units as the one dimensional case. These and subsequent plots are scaled in the y-direction so that the complete set may be displayed together, for ease of comparison. }
\label{clumpy}
\end{figure*}

\subsection{Velocity perturbed flow}
\begin{figure*}
\caption{The evolution of a velocity perturbation of 20 per cent above the
background level is shown at variety of times, labelled beneath each
plot. Many interesting features can be seen in this 2D model. As the
perturbed gas moves through the background it appears to trigger
Richtmeyer-Meshkov instabilities in the dense shells and causes a
disturbance to the upstream background.}
\label{2d_wide1}
\end{figure*}

\label{2d_velo}
As a first example of the development of a perturbed two dimensional
flow, we set the flow velocity to be $20$ per cent greater than its
value at the equivalent depth in the surrounding flow in a region of
$10 \times 40$ grid elements (at a resolution of $1$ element $=
0.00375$) at the base of the wind.  Initially the disturbance expands
smoothly through the background, and no significant effects appear on
large scales.  The matter in the initially perturbed region collects
into a dense clump, with rarefied regions immediately upstream and
downstream of it.  The downstream rarefaction is simply the result of
the increased velocity of the gas in the perturbed region, while the
upstream perturbation is caused by the shadow of the perturbation.  In
this shadow, the gas experiences a decreased radiation force and so
slows and is swept up by the clump.  

As the perturbation increases in size, the changes of shadowing from
gas shadowed by the perturbation (etc.) begin to have effects on the
dense shells far downstream.  The patch of the dense shell marked (i)
in Fig.~\ref{2d_wide1}b, falls in the shadow of the perturbation,
decelerates and becomes more diffuse, causing the first break in a
shell structure.  At this time an instability develops in the
unshadowed parts of the dense shell where the optically thin faster
flowing gas meets an optically thick slower dense shock.  The
disruption of the shells is dependent on whether they continue to
contract as they evolve, which is in turn dependent on whether they
are compressed by a shock behind or retarded by a dense region in
ahead.  If the shell expands slower than the perturbation evolves, the
RM instability within the shell structure develops.  If not the
unstable structure diffuses away as the gas expands.  This is the case
in shell (ii) of Fig.~\ref{2d_wide1}c where the increasing radial
width and decreasing density of the shell damps the instability, so
that before it becomes significant the shell disperses.  The shadowed
region is unaffected by this expansion due to the alterations in
spectrum that govern its motion.  The apex of the shadowed region in
the same plot forms a transient dense feature labelled (iii) which is
created by the fast highly rarefied region immediately upstream of it,
from which it accretes gas.  As the flow evolves further, this clump
disperses.

It is interesting to note that when part of a dense shell falls into
the rarefied slipstream of the perturbation it experiences very little
driving and is swept up by the dense clump of the perturbation. There
is steady accretion from such incidents, increasing the density and
extent of the perturbation. Were it not for this accretion the
perturbation would become highly rarefied and disappear, but as is
shown in Fig.~\ref{2d_wide1}d, part of the dense shell is accelerated
and compressed into a high density bullet which rams through the
perturbation.

Further downstream, Fig.~\ref{2d_wide1}e this feature develops into a
bow shock/arrow shape and added to further by faster flowing gas
accreting onto it from behind.  This structure is preserved until the
edge of the grid, but it is unlikely to remain in this form in the far
downstream flow.  When the wind has reached the terminal velocity
adiabatic expansion will cause it to diffuse away.

Persistent features appear in the inner wind at the position of the
initial perturbation.  Instabilities seeded by these disrupt the
formation of dense shells in the downstream flow.  This is eventually
smoothed out as the features spread in the tangential direction,
Fig.~\ref{2d_wide1}f.  However, as the flow is everywhere supersonic,
the perturbations ought to have been advected downstream long before.

In an Eulerian hydrodynamical scheme, perturbations to smooth flow in
a particular cell can only be removed exponentially with time (in
essence by dilution).  A sufficiently strong instability can decrease
the exponential damping rate, or even force the perturbations to grow.
This may be viewed as a limitation to the accuracy of these
simulations -- as indeed it would be a limitation to the accuracy of
{\it any}\/ Eulerian simulation with finite resolution (Lagrangian
simulations would suffer from complementary problems as a result of
sampling noise).  However, in reality the flow will continually be
affected by small scale perturbations from the photospheric flow.  At
late times, as the dense but noisy shells evolve they become highly
structured as the hydrodynamics and the line-driving struggles for
dominance in these regions.  This flow is similar to that for
unperturbed initial conditions, which suggests that it is not
sensitive to the precise nature of the upstream perturbations, so long
as they are small -- the exponential growth of the instability erases
the small details of its seeding.

\subsection{Density perturbed flow}
\begin{figure*}
\caption{The evolution of a density perturbation of 20 per cent above 
the background level is shown at variety of times, labelled beneath
each plot. These results contrast with those for a velocity
perturbation shown earlier. Although the flow is similar the direction
of the perturbation fronts is reversed. }
\label{2d_wide2}
\end{figure*}

Contrasting behaviour is observed if a density perturbation is applied
to the background flow.  The motivation for this choice of
perturbation was interest in a macroscopic persistent feature in the
outer portion of the wind.  The phenomenon of Co-rotating Interaction
Regions (CIRs) has been discussed previously (e.g. Dessart \& Chesneau
2002) but as yet their formation is not well understood.

Initially, the perturbation expands in the form expected for such an
overpressured region from pure hydrodynamics.  A diffuse bubble
appears, with dense shocked plates at its upstream and downstream
edges.  However, as the flow develops, these plates move apart and
become more condensed, while the material between them is accelerated
by line driving into the downstream clump, or is swept up by the
approach of the upstream edge or clump.  The direction which material
will move and which clump it will be swept into is dependent on
whether the gas falls in the shadow of the upstream clump or not.

The downstream flow still shows evidence of the spectral interference
from the perturbation.  The background dense shells are disturbed
initially on the line of the edges of the perturbation, since at these
positions a range of velocities are present while in the body of the
perturbation has a very narrow velocity spread; the downstream
material is only disturbed if gas upstream of it blocks radiation in a
relevant frequency range.

The subsequent evolution of the upstream and downstream components of
the original perturbation is distinct from that found in the previous
section, showing sensitive dependence of the details of the flow on
the initial conditions.  One obvious feature, shown in
Fig.~\ref{2d_wide2}b, is that both clumps of the perturbation appear
to be blown away by the outflowing wind whereas in
Fig.~\ref{2d_wide1}b the perturbation forms bow shocks as it pushes
through slower material ahead of it.  

These results prompted an investigation of a higher density initial
perturbation, so a density perturbation by a factor of 4 times was
applied to the same initial conditions as above.  As one might expect
the evolution of this perturbation, as seen in Fig.~\ref{2d_wide3}a-f
is similar to that seen in Fig.~\ref{2d_wide2}a-f.  In this case the
upstream edge of the expanding bubble withstands disruption by the
flow and eventually forms a 'v'-shaped front, which after further
evolution develops highly detailed structure in Fig.~\ref{2d_wide3}d
as a shock passes through it leaving RM instabilities to develop.

The structure from the downstream edge of the bubble forms features in
Fig.~\ref{2d_wide3}c-d that are reminiscent of the structure in
Fig.~\ref{2d_wide2}c-d. The perturbation has a great effect on the
persistent features which remain in the upstream flow, leading to the
beginning of highly clumped flow in Fig.~\ref{2d_wide3}f.

\begin{figure*}
\caption{The evolution of a density perturbation a factor of 4 times the background level is shown at variety of times, labelled beneath each plot. }
\label{2d_wide3}
\end{figure*}

\subsection{Physical consequences}

Discrete Absorption Components \cite[DACs; e.g.]{kaphen99} are
significant features seen in the absorption profiles of unsaturated
lines.  Current observation suggests that the kinematics of these
features is dominated by a rotational component, but theory has not
addressed the manner in which they form and their dynamics as they
move through the wind.  The results of the above sections suggest that
features that are formed in the early parts of the wind, are dispersed
rapidly if their propagation is slow enough to be destroyed by RM
instabilities.  DACs are formed in the early part of a real wind when
a slowly rotating region suffers a compression from a neighbouring
fast moving region.  Higher density compressed portions of the wind
are observed in line spectra as DACs.

The dense, long lived bullets which are produced in our simulations
are a strikingly similar to those inferred in some empirical models of
DACs \cite[e.g.]{hambr01}.  Similar structures would be expected to
form as a result of variations in the stellar flux due to dark surface
features, which may naturally explain the origin of CIRs.

The ultimate aim of these models is to understand the nature of
observed hot star winds.  While the present results are by no means
definitive, it is clear that cohesive absorbing knots are a natural
feature of line-driven winds.  However, further, more physically
detailed, numerical simulations will be required before a detailed
comparison to the observations of wind structures such as DACs can be
performed.

\section{Summary}

We present models of the structure of unstable line-driven flows,
calculated in two dimensions with a full hydrodynamical code and
non-local radiation transport.

Our results compare well to those of more physically complete (but
geometrically restricted) models in their common limits.  The
reproduction of the noise waves and the dominance of reverse shocks in
the downstream structure is in good agreement with those seen by other
non-Sobolev approaches (e.g. OCR, Runacres \& Owocki 2002, Feldmeier
1995).  This behaviour is borne out by the close similarity of the
statistical descriptors of the flows we study to the results presented
by Runacres \& Owocki (2002).  This suggests that the current model
does indeed present a good description of the non-Sobolev behaviour of
this wind.

The model presented here has limitations: 1) the gas is taken to be
isothermal, which is reasonable at least in high density regions; 2)
the spatial domain is restricted, effectively to a small region
reasonably far from the photosphere; 3) the line profile function is
taken to have a top-hat form rather than a Gaussian or Voigt profile;
4) the fiducial value of $v|{th}/c|s = 0.5$ was
used~\cite[following]{owoc91}, although a rather lower value may in
fact be more appropriate; 5) we have treated the radiation transfer in
the pure-absorption approximation.  We have argued, however, that none
of these approximations will have a significant effect on the local
structure of the line-driven winds, which is our principal interest in
this paper: our work complements other studies in which these
processes are treated in more physical detail in a more geometrically
restricted domain.

In future work, we will include the line-drag effect of scattered
radiation~\cite{lucy84} \cite[formulated in the non-Sobolev method
by]{owocp96}.  The asymmetry we have shown between forward and
backward calculation of the Sobolev line force from the wind velocity
gradient, could be of use in approximating the fore/aft asymmetry
which produces line drag.  Since this is a local effect, it may not be
necessary to use a full non-Sobolev method for its calculation.
Indeed a Sobolev approach may actually be more suitable than a global
method for simulations with high degrees of imposed symmetry.

\section*{Acknowledgments}

This work was supported by PPARC through a research studentship (ELG)
and an Advanced Fellowship (RJRW), and by NASA through grant
NAG5-12020.  We would like to thank Mark Runacres and Ian Stevens for
several helpful discussions, and the referee, Stan Owocki, for a
constructive and helpful report.


\end{document}